%% file: conference_101719.tex
\documentclass[conference]{IEEEtran}
\usepackage{cite}
\usepackage{booktabs}
\usepackage{multirow}
\usepackage{amsmath,amssymb,amsfonts}
\usepackage{tabularx}
\usepackage{siunitx}
\usepackage{algorithm}
\usepackage{algpseudocode}
\usepackage{graphicx}
\usepackage{textcomp}
\usepackage[square,sort,comma,numbers]{natbib}
\usepackage{xcolor}
\def\BibTeX{{\rm B\kern-.05em{\sc i\kern-.025em b}\kern-.08em
    T\kern-.1667em\lower.7ex\hbox{E}\kern-.125emX}}

\begin{document}

\title{Enhancing Clinical Trial Patient Matching through Knowledge Augmentation and Reasoning with Multi-Agent}

\author{\IEEEauthorblockN{1\textsuperscript{st} Hanwen Shi}
\IEEEauthorblockA{\textit{DO\&IT Department} \\
\textit{University of Maryland}\\
College Park, USA \\
hwshi@umd.edu}
\and
\IEEEauthorblockN{2\textsuperscript{nd} Jin Zhang}
\IEEEauthorblockA{\textit{Independent Researcher}\\
Ellicott City, USA \\
zhangjin1980@hotmail.com}
\and
\IEEEauthorblockN{3\textsuperscript{st} Kunpeng Zhang}
\IEEEauthorblockA{\textit{DO\&IT Department} \\
\textit{University of Maryland}\\
College Park, USA \\
kpzhang@umd.edu}
}

\maketitle

\begin{abstract}
Matching patients effectively and efficiently for clinical trials is a significant challenge due to the complexity and variability of patient profiles and trial criteria. This paper introduces \textbf{Multi-Agent for Knowledge Augmentation and Reasoning (MAKAR)}, a novel multi-agent system that enhances patient-trial matching by integrating criterion augmentation with structured reasoning. MAKAR consistently improves performance by an average of 7\% across different datasets. Furthermore, it enables privacy-preserving deployment and maintains competitive performance when using smaller open-source models. Overall, MAKAR can contributes to more transparent, accurate, and privacy-conscious AI-driven patient matching.
\end{abstract}


\section{Introduction}

Patient matching for clinical trials is the process of identifying and enrolling participants whose health profiles meet the specific eligibility criteria of a given clinical study. Efficient patient matching is crucial to accelerating the drug development process and reducing the overall cost of clinical trials. It also provides an opportunity for participants to access experimental treatments that could potentially improve their health outcomes and, in some cases, transform their quality of life.

However, patient matching remains a challenging task despite its critical role in clinical trials. Traditional automated methods, which are mostly rule-based, depend heavily on hand-crafted rules and are often time-consuming~\citep{brogger2020online}. These approaches are also highly sensitive to variations in how patient and trial data are expressed in natural language across different sources. Simple machine learning models often struggle with the load of information and the low density of relevant sentences in lengthy electronic health record(EHR) data. They usually underperform when dealing with complex eligibility criteria or inconsistent data formats, resulting in inefficiencies and mismatches that can negatively impact trial outcomes~\citep{kadam2016challenges}.

The emergence of  of LLMs offers a promising opportunity to improve patient matching. LLMs, such as GPT-4~\cite{achiam2023gpt}, usually possess a large text window and advanced natural language understanding capabilities~\citep{karanikolas2023large}, which enable them process complex trial criteria and patient data effectively. However, three limitations constrain the performance of LLMs. First, \textit{knowledge gaps}~\citep{shuster2021retrieval} may exist due to training data cut-off dates and incomplete coverage of specialized medical knowledge, such as newly approved drugs, treatments, or updated FDA guidelines. Second, \textit{hallucinations}~\citep{huang2025survey} remain a significant challenge: LLMs are often prompted to generate detailed reasoning to improve interpretability and trustworthiness in patient matching~\citep{jin2024matching}, yet such reasoning is unverified and potentially misleading. Finally, \textit{prompt sensitivity} can substantially impact outputs, as even minor variations in prompts can lead to divergent results~\citep{zhuo-etal-2024-prosa}.

To address these limitations, we present a training-free multi-agent workflow: Multi-Agent for Knowledge Augmentation and Reasoning (MAKAR). MAKAR consists of two multi-agent modules: the Augmentation Module and the Reasoning Module. The augmentation module enriches trial criteria by providing detailed explanations from different sources. And the reasoning module evaluates each criterion-related patient condition in a step-by-step manner to determine eligibility and make the final matching decision. We demonstrate the effectiveness of MAKAR our framework on public dataset and in-house real-world database. On average, MAKAR achieves an improvement of 7\% over benchmark methods, with up to a 10\% improvement on certain criteria. Our key contributions are as follows:
\begin{itemize}\setlength\itemsep{0.2em}
\item We introduce MAKAR, a double-module multi-agent workflow that allows complementary contributions from criterion augmentation and reasoning. And it achieves consistent performance improvement across different datasets.
\item We demonstrate that MAKAR remains effective when deployed with open-source language models. And it has potential to contribute to more trustworthy and privacy-preserving AI application in drug development.
\end{itemize}

\begin{figure*}
    \centering
    \includegraphics[width=\textwidth]{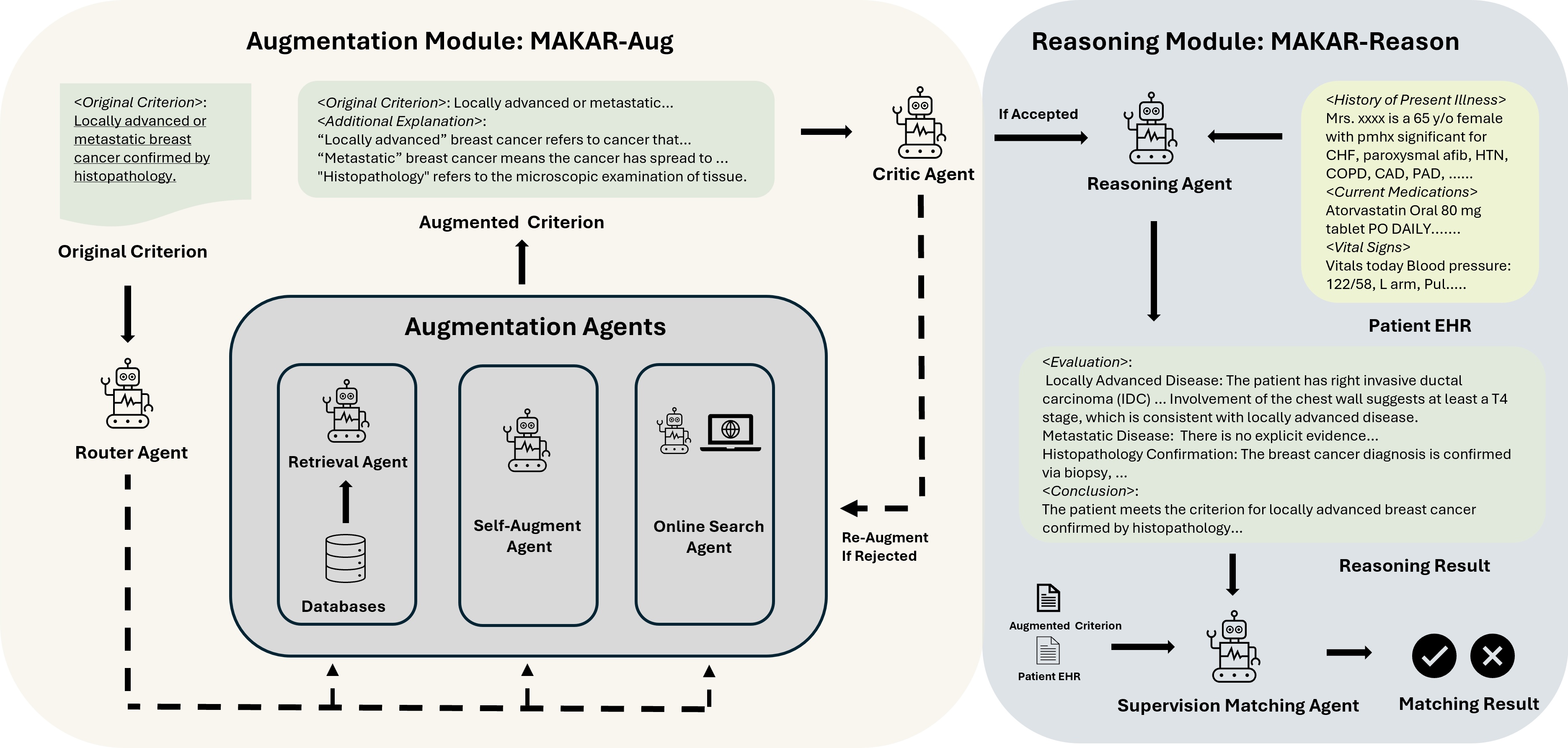}
    \caption{The workflow of MAKAR framework}
    \label{fig:framework}
\end{figure*}

\section{Related Work}
This work falls within the literature on patient matching using free-text patient records. Early approaches such as RBC~\cite{oleynik2019evaluating} relied on rule-based classification. Transformer encoder-based methods were later introduced, including DeepEnroll~\cite{zhang2020deepenroll} and COMPOSE~\cite{gao2020compose}, which improved the ability to encode complex eligibility criteria and patient notes. More recently, generative large language models (GLLMs) have been applied to this task and have shown stronger performance, as in TrialGPT~\cite{jin2024matching}, CoT matching~\cite{beattie2024utilizing}, and other zero-shot methods~\cite{wornow2024zero}. In parallel, related work~\cite{loaiza2025transforming} explored the use of multi-agent systems to analyze oncology knowledge graphs for patient enrollment.

Another related line of research is LLM-based multi-agent systems (LLM-MA). In the healthcare domain, applications of LLM-MA are rapidly gaining traction. Yue et al.\cite{yue2024ct} introduce ClinicalAgent, a multi-agent architecture powered by LLM reasoning for multi-agent collaboration, achieving higher precision in predicting clinical trial outcomes. Similarly, Liao et al.\cite{liao2024reflectool} propose REFLECTool, a multi-agent framework that enables agents to learn how to select domain-specific tools, thereby extend their capabilities. TriageAgent\cite{lu2024triageagent} leverages LLMs for role-playing interactions, integrating self-confidence assessments and early-stopping mechanisms in multi-round discussions to improve document reasoning and classification accuracy for triage tasks. Tang et al.\cite{tang2023medagents} use multiple LLM-based agents as domain experts to collaboratively analyze medical reports. And Li et al.~\cite{li2024mmedagent} further introduce the multi-agent collaboration framework to handle multimodal tasks.

\input{table/main_result}
\section{Methods}
Our proposed methods, shown in Figure~\ref{fig:framework}, is a training-free multi-agent workflow and consists of two core modules: the \textit{Augmentation Module} and the \textit{Reasoning Module}. The augmentation module is designed to generate criteria with comprehensive explanations of relevant concepts, while the reasoning module evaluates each condition in the criteria to determine eligibility and make the final matching decision. The workflow uses five different types of agent and the prompts used by the agents can be found in Appendix~\ref{app:prompt_agent}. The pseudo code of the workflow is provided in Appendix~\ref{app:pseudio}

\textbf{Router Agent} 
In MAKAR, multiple specialized augmentation agents are available to augment the criterion with domain-specific knowledge. And the Router Agent determines the most suitable augmentation agent for the task and routes the workflow accordingly. Inspired by SELFREF~\cite{kadavath2022language} and MOREINFO~\cite{feng2023knowledge}, we adopt a prompt-based self-probing strategy. If the agent determines that the original criterion is detailed and understandable enough for other agents, the criterion will be directly forwarded to the Reasoning Module. Otherwise, the criterion is forwarded to the Knowledge Augmentation Agent chosen by the Router Agent for further processing.

\textbf{Knowledge Augmentation Agent:}  
This component augments the criterion with relevant, domain-specific information from different source. For instance, the \textit{Retrieval Agent}~\cite{lewis2020retrieval} retrieves information from indexed medical databases, the \textit{Online Search Agent} gathers supplementary knowledge from online sources, and the \textit{Self-Augment Agent} leverages the LLM’s intrinsic knowledge to refine the original criterion. 

\textbf{Critic Agent:}  
We add Critic Agent as the supervisor to power the iterative refinement with self-feedback as in SELF-REFINE~\cite{madaan2023self}. It ensures that the augmented criterion is aligned with the original instructions. For instance, the original criterion must be preserved verbatim. Additionally, the critic agent verifies that the augmented output is well-structured (e.g., concepts are clearly itemized) to facilitate further step-wise reasoning. This critic mechanism are designed to improve the framework's ability to follow instructions accurately while maintaining consistency and clarity.

\textbf{Reasoning Agent}
With the augmented criterion and patient EHR data, the Reasoning Agent evaluates patient eligibility by assessing for each concepts mentioned in the augmented criterion. It generates a detailed reasoning summary~\cite{wornow2024zero}, including assessments for all relevant health conditions, and forwards it to the Matching Agent.

\textbf{Matching Agent:}  
The Matching Agent reviews the reasoning assessment alongside the patient’s EHR data and the augmented criterion. Then it determines patient eligibility in a zero-shot manner.

\section{Experiments}
\subsection{Data}
We use a widely-adopted public dataset and an in-house real-world dataset to evaluate our proposed method. Both datasets were labeled by at least two trained annotators, with any disagreements resolved by an independent physician to determine the final ground truth. The public dataset is from Track 1 of the 2018 n2c2 challenge~\cite{stubbs2019cohort}. The dataset consists of 288 de-identified patients, each having between 2 to 5 unstructured longitudinal clinical notes in American English. The challenge simulates a synthetic clinical trial with 13 predefined inclusion criteria (see Appendix~\ref{sec:original_cri}), resulting in 3744 patient-criteria pairs. Each pair has a binary label—“MET” or “NOT MET” and the statistics can be found in Table~\ref{tab:met_counts}.

To assess our method in a more complex setting, we created a real-world dataset by sampling 30 eligibility criteria (15 inclusion, 15 exclusion) from ongoing breast cancer trials on ClinicalTrials.gov. We manually selected 10 breast cancer patients from our in-house EHR system and extracted their most recent deidentified clinical records. When information was insufficient to confirm eligibility, the criterion was conservatively labeled as "NOT MET." The sampled criteria are listed in Appendix~\ref{sec:clinicalgov_cri}.
\subsection{Baselines and Metrics}
Previous research has shown that LLM-based approaches outperform traditional rule-based systems~\cite{jin2024matching}. Thus, this study focuses exclusively on LLM-based benchmarks. We include three benchmarks: (1) TrialGPT~\cite{jin2024matching}, (2) the two-stage retrieval-based zero-shot matching system~\cite{wornow2024zero}, and (3) the chain-of-thought (CoT) based method~\cite{beattie2024utilizing}. For the n2c2 dataset, we report accuracy and F1-score for each criterion as well as the overall average. For the ClinicalTrial dataset, we report the average accuracy and F1-score for inclusion and exclusion criteria separately.
\subsection{Experiment Settings}
Our experiments were conducted on 4 A10G GPU with 96 GB memory. Unless otherwise specified, all agents in the framework are powered by GPT-4o.

For external knowledge augmentation, the Online Search Agent uses the Perplexity AI search engine to fetch up-to-date information from the web. The Retrieval Agent uses all-MiniLM-L12-v2, a pre-trained checkpoint from Sentence-Transformer~\cite{reimers2019sentence}, as its encoder to construct the vector database of MedDRA and FDALabel.

\section{Results}
Table~\ref{tab:main_result} compares the performance of MAKAR with benchmark methods across two datasets. Overall, the MAKAR model consistently achieves the highest or near-highest accuracy and F1-scores across most conditions. Approximately, MAKAR improves accuracy by 4\% and F1-score by 7\% on average. And it shows particularly strong performance gains (more than 10\%) on criteria such as CREATININE, HBA1C, and MI-6MOS.

To better understand these improvements, we examine the MAKAR-augmented criteria presented in Appendix~\ref{sec:gpt_cri}. Compared to the baseline human-refined criteria outlined in Appendix~\ref{sec:human_cri}, MAKAR has two major improvements: (1) more detailed expansions of key concepts, for example, supplementing normal ranges for criteria like HBA1C and CREATININE across diverse demographic groups; and (2) more comprehensive and accessible definitions, such as using widely recognized terms like “heart attack” alongside the clinical term “myocardial infarction (MI)”. These refinements help LLMs to match patients from clinical notes written in various styles and lead to substantial performance gains on these criteria.

Patient matching is not merely a binary classification task; the correctness of the reasoning process is critical for transparency, which in turn affects physicians’ trust in AI systems~\cite{asan2020artificial}. To evaluate it, we manually reviewed the reasoning outputs for all 300 patient–criterion pairs. Although MAKAR produced entirely correct final classifications on the ClinicalTrial dataset, we identified two minor flaws in the intermediate reasoning steps. And the results reported in Table~\ref{tab:main_result} reflect the quality of both the reasoning process and the final matching outcomes.

First, in Exclusion Criterion 13 (\textit{Patients with ongoing indications for cardioprotective medication—ACE inhibitors, ARBs, and/or beta-blockers}), MAKAR reasoned: \textit{"The patient is on metoprolol, a beta-blocker, suggesting some ongoing indication for its use. However, without explicit information on the specific condition necessitating this medication (e.g., arrhythmia, prophylactic use post-cardiac event other than MI, or another reason), the justification remains uncertain."}. Here, the Reasoning Agent is overly conservative, but the Supervision Matching Agent correctly classified this case as MET. This case shows the design is effective.

Second,in Inclusion Criterion 5 (\textit{Histologically confirmed breast cancer with an invasive component $\ge$ 20 mm and/or morphologically confirmed spread to regional lymph nodes(stage cT2-cT4 with any cN, or cN1-cN3 with any cT).}), MAKAR wrongly inferred that a 1.5 cm lesion is greater than 20 mm. However, because lymph node involvement was documented, the final decision remained correct.

\input{table/ablation1}

\section{Ablation}
\subsection{Augmentation and Reasoning}
In this section, we validate the double-module design of our workflow on ClinicalTrial dataset. First, we remove the knowledge augmentation module to assess the reasoning module alone (MAKAR-Reason, equivalent to a CoT) using the original criteria described in Appendix~\ref{sec:original_cri}. Next, we remove the reasoning module and evaluate the augmentation module (MAKAR-Aug) using zero-shot matching with prompts detailed in Appendix~\ref{app:prompt_zero}). The results, presented in Table~\ref{tab:ablation1}, suggest that both modules contributes to overall performance. Specifically, removing the augmentation module (MAKAR-Reason alone) led to a 6\% drop in F1 score, while removing the reasoning module (MAKAR-Aug alone) led to a 5\% drop.

To further demonstrate the effectiveness of the reasoning module, we compared its performance against other reasoning models. Interestingly, MAKAR-Reason outperformed both OpenAI o1 and o3 mini, which can be attributed to the chain-of-thought strategy used by the reasoning agent. For example, when evaluating whether a patient met Inclusion Criterion 14 \textit{(Postmenopausal hormone receptor-positive patients)}, MAKAR reasoned: \textit{"...the record does not explicitly state whether the breast cancer was hormone receptor positive. However, the use of tamoxifen, which is specifically prescribed for estrogen receptor-positive breast cancer by blocking estrogen receptors, strongly implies that her breast cancer was hormone receptor positive."} Both MAKAR and MAKAR-Reason correctly identified this case, while MAKAR-Aug failed to capture the implicit relationship.

Additionally, we also observed cases where criterion augmentation improved performance. For example, in Exclusion Criterion 7, it mentions: \textit{"...all histological lesions were HER2 1+ or 2+, not detected by FISH amplification, and HR positive according to ASCO guidelines"}. MAKA-aug expanded the definitions from American Society of Clinical Oncology(ASCO) guidelines, made correct matches for all 10 patients. In contrast, MAKA-reason only made 8 correct matches out of 10.

\input{table/ablation2}

\subsection{Local Deployment}
One key challenge for applying MAKAR in real-world settings is that agents in the reasoning module need to process patient EHR data directly. And such data are extremely sensitive and protected under the Health Insurance Portability and Accountability Act (HIPAA). It requires de-identifying large volumes of patient information before uploading data to a closed-source external model like GPT-4o. And this process can be time-consuming and still has residual privacy risks. To address this limitation, we assessed whether MAKAR can maintain good performance after replacing GPT-4o with open-source language models that can be deployed locally. 

Our local deployment experiments consist of three configurations: (1)Zero-Shot Matching, where we applied open-source models directly in zero-shot mode using the criteria described in Appendix~\ref{sec:clinicalgov_cri} and the prompts from Appendix~\ref{app:prompt_zero}; (2) GPT-Aug, where we used GPT-4 to augment the criteria (since criteria text itself is not sensitive) and then applied a open-source model with the same zero-shot prompt; and (3) MAKAR, where we replaced all agents in the reasoning module with open-source models

The experiments were conducted on the ClinicalTrial dataset, and accuracies are reported in Table~\ref{tab:ablation2}. We found that replacing GPT-4o in the reasoning module with locally deployed open-source models reduced performance by an average of 11\%. However, some configurations still have competitive performance; for example, MAKAR powered by the Qwen family achieved accuracy greater than 0.900. The results further confirm the effectiveness of the design of MAKAR and it improved performance by approximately 5\% compared to zero-shot open-source benchmarks. 

Another interesting observation is that smaller models appeared less sensitive to augmented criteria: models under 7B parameters showed no performance gains from augmentation. In contrast, both smaller and larger models consistently benefited from the reasoning module, with accuracy improving after reasoning module was applied. However, this observation should be interpreted with caution, as it is based on a limited number of model configurations evaluated in this study.

\section{Conclusion}
In this study, we introduced MAKAR, a multi-agent workflow designed to enhance patient-trial matching by combining criterion augmentation with structured reasoning. Our experimental results show that MAKAR achieves an average improvement of 7\% and up to 10\% on certain criteria. Through detailed ablation studies, we further demonstrate that both modules of the MAKAR design contribute independently to these gains. In addition, our local deployment experiments confirmed that MAKAR remains effective when using open-source language models. Even with smaller models, such as Qwen and Gemma, the MAKAR workflow achieved competitive performance. These findings suggest that MAKAR has the potential to enable more accurate and privacy-preserving patient selection.

\section*{Limitations and Discussion}
One observation from our experiments is that the Router Agent consistently determined that all criteria could be asugmented by the Self-Augmentation Agent, without requiring external retrieval or online search augmentation. This was true even for breast cancer-related criteria, which are typically complex and challenging for individuals without professional training.

This relates to a limitation of our study, which comes from the dataset used for evaluation. Both the n2c2 dataset and the real-world dataset are relatively small and contain criteria that are not particularly challenging for LLM-based processing, allowing all benchmark methods to achieve strong performance. While MAKAR is designed as a general framework capable of handling both simple and complex criteria, these dataset constraints limit our ability to explore the knowledge boundaries of LLMs and fully assess our method's generalizability across a wider range of clinical trial requirements.

However, despite the dataset limitations, the effectiveness of the augmentation module remains significant. One of its key strengths is its potential to interface with diverse data sources, particularly those that are not publicly available, such as paid subscription-based medical databases or proprietary clinical trial repositories. Furthermore, it plays a crucial role in integrating the most up-to-date medical knowledge, including newly developed treatments and evolving FDA guidelines. This capability ensures that the framework remains adaptable and relevant, even when dealing with rapidly changing medical landscapes where LLMs alone may lack the latest information. By systematically incorporating external knowledge, the augmentation module enhances the contextual relevance of trial matching, making it particularly valuable in real-world applications where timely and comprehensive knowledge retrieval is essential.

\bibliographystyle{IEEEtran}
\bibliography{custom}

\newpage

\appendix

\section{Statistics}
\input{table/stats}
\section{Criteria}
\subsection{N2C2 Original Criteria}\label{sec:original_cri}
\noindent\textbf{DRUG-ABUSE}: Drug abuse, current or past.

\noindent\textbf{ALCOHOL-ABUSE}: Current alcohol use over weekly recommended limits.

\noindent\textbf{ENGLISH}: Patient must speak English.

\noindent\textbf{MAKES-DECISIONS}: Patient must make their own medical decisions.

\noindent\textbf{ABDOMINAL}: History of intra-abdominal surgery, small or large intestine resection, or small bowel obstruction.

\noindent\textbf{MAJOR-DIABETES}: Major diabetes-related complications, such as: amputation, kidney damage, skin conditions, retinopathy, nephropathy, neuropathy.

\noindent\textbf{ADVANCED-CAD}: Advanced cardiovascular disease, defined as two or more of the following: medications for CAD, history of myocardial infarction, angina, ischemia.

\noindent\textbf{MI-6MOS}: Myocardial infarction in the past 6 months.

\noindent\textbf{KETO-1YR}: Diagnosis of ketoacidosis in the past year.

\noindent\textbf{DIETSUPP-2MOS}: Taken a dietary supplement (excluding Vitamin D) in the past 2 months.

\noindent\textbf{ASP-FOR-MI}: Use of aspirin to prevent myocardial infarction.

\noindent\textbf{HBA1C}: Any HbA1c value between 6.5 and 9.5\%.

\noindent\textbf{CREATININE}: Serum creatinine > upper limit of normal.

\subsection{N2C2 Criteria Refined by \cite{wornow2024zero}}\label{sec:human_cri}
\noindent\textbf{ABDOMINAL}: History of intra-abdominal surgery. This could include any form of intra-abdominal surgery, including but not limited to small/large intestine resection or small bowel obstruction.

\noindent\textbf{ADVANCED-CAD}: Advanced cardiovascular disease (CAD). For the purposes of this annotation, we define “advanced” as having 2 or more of the following: (a) Taking 2 or more medications to treat CAD (b) History of myocardial infarction (MI) (c) Currently experiencing angina (d) Ischemia, past or present. The patient must have at least 2 of these categories (a,b,c,d) to meet this criterion, otherwise the patient does not meet this criterion. For ADVANCED-CAD, be strict in your evaluation of the patient—if they just have cardiovascular disease, then they do not meet this criterion.

\noindent\textbf{ALCOHOL-ABUSE}: Current alcohol use over weekly recommended limits.

\noindent\textbf{ASP-FOR-MI}: Use of aspirin for preventing myocardial infarction (MI).

\noindent\textbf{CREATININE}: Serum creatinine level above the upper normal limit.

\noindent\textbf{DIETSUPP-2MOS}: Consumption of a dietary supplement (excluding vitamin D) in the past 2 months. To assess this criterion, go through the list of medications and supplements taken from the note. If a substance could potentially be used as a dietary supplement (i.e., it is commonly used as a dietary supplement, even if it is not explicitly stated as being used as a dietary supplement), then the patient meets this criterion. Be lenient and broad in what is considered a dietary supplement. For example, a “multivitamin” and “calcium carbonate” should always be considered a dietary supplement if they are included in this list.

\noindent\textbf{DRUG-ABUSE}: Current or past history of drug abuse.

\noindent\textbf{ENGLISH}: Patient speaks English. Assume that the patient speaks English, unless otherwise explicitly noted. If the patient’s language is not mentioned in the note, then assume they speak English and thus meet this criterion.

\noindent\textbf{HBA1C}: Any hemoglobin A1c (HbA1c) value between 6.5\% and 9.5\%.

\noindent\textbf{KETO-1YR}: Diagnosis of ketoacidosis within the past year.

\noindent\textbf{MAJOR-DIABETES}: Major diabetes-related complication. Examples of “major complication” (as opposed to “minor complication”) include, but are not limited to, any of the following that are a result of (or strongly correlated with) uncontrolled diabetes: Amputation, kidney damage, skin conditions, retinopathy, nephropathy, neuropathy. Additionally, if multiple conditions together imply a severe case of diabetes, then count that as a major complication.

\noindent\textbf{MAKES-DECISIONS}: Patient must make their own medical decisions. Assume that the patient makes their own medical decisions, unless otherwise explicitly noted. If there is no information provided about the patient’s ability to make their own medical decisions, then assume they do and therefore meet this criterion.

\noindent\textbf{MI-6MOS}: Myocardial infarction (MI) within the past 6 months.

\subsection{N2C2 Criteria Refined by GPT-4o}\label{sec:gpt_cri}
\noindent\textbf{ABDOMINAL}

\noindent\textit{Criteria:} History of intra-abdominal surgery, including small/large intestine resection or small bowel obstruction.

\noindent\textit{Explanation:} Intra-abdominal surgery involves operations on organs within the abdominal cavity (e.g., stomach, intestines). Small bowel obstruction refers to blockages in the small intestine that may require surgery.

\noindent\textbf{ADVANCED-CAD}

\noindent\textit{Criteria:} Advanced cardiovascular disease (CAD) defined by having two or more of the following: (a) Taking 2 or more medications for CAD, (b) History of myocardial infarction (MI), (c) Current angina, or (d) Ischemia (past or present).

\noindent\textit{Explanation:} CAD affects heart function due to blocked arteries. Definitions include: MI (heart attack), angina (chest pain), ischemia (reduced blood flow), and specific medications like statins or beta-blockers. A diagnosis of CAD alone does not qualify as advanced.

\noindent\textbf{ALCOHOL-ABUSE}

\noindent\textit{Criteria:} Current alcohol use exceeding weekly recommended limits.

\noindent\textit{Explanation:} Weekly limits: up to 14 standard drinks. A standard drink is approximately 14g of pure alcohol (12oz beer, 5oz wine, or 1.5oz spirits).

\noindent\textbf{ASP-FOR-MI}

\noindent\textit{Criteria:} Use of aspirin to prevent myocardial infarction (MI).

\noindent\textit{Explanation:} Aspirin reduces MI risk by preventing clots but requires medical advice to balance benefits (risk reduction) and potential side effects (e.g., bleeding).

\noindent\textbf{CREATININE}

\noindent\textit{Criteria:} Serum creatinine level above the upper normal limit.

\noindent\textit{Explanation:} Creatinine reflects kidney function. Normal levels vary by age and gender (e.g., approximately 0.6--1.2 mg/dL for adult males). Elevated levels suggest potential kidney impairment.

\noindent\textbf{DIETSUPP-2MOS}

\noindent\textit{Criteria:} Consumption of a dietary supplement (excluding Vitamin D) in the past 2 months.

\noindent\textit{Explanation:} Supplements include multivitamins, calcium carbonate, fish oil, or herbal supplements like ginseng. Excludes Vitamin D explicitly.

\noindent\textbf{DRUG-ABUSE}

\noindent\textit{Criteria:} Current or past drug abuse history.

\noindent\textit{Explanation:} Drug abuse includes misuse of psychoactive substances, with potential consequences like health issues or legal problems.

\noindent\textbf{ENGLISH}

\noindent\textit{Criteria:} Patient must speak English.

\noindent\textit{Explanation:} English proficiency ensures effective participation in English-based trials. Assume English proficiency unless noted otherwise.

\noindent\textbf{HBA1C}

\noindent\textit{Criteria:} Hemoglobin A1c (HbA1c) between 6.5\% and 9.5\%.

\noindent\textit{Explanation:} HbA1c measures average blood sugar over 2--3 months. The range identifies diabetes patients with moderate control, excluding poorly controlled cases.

\noindent\textbf{KETO-1YR}

\noindent\textit{Criteria:} Diagnosis of ketoacidosis in the past year.

\noindent\textit{Explanation:} Ketoacidosis is a serious condition with high blood ketones, often related to uncontrolled diabetes. Symptoms include nausea, confusion, and rapid breathing.

\noindent\textbf{MAJOR-DIABETES}

\noindent\textit{Criteria:} Major diabetes-related complications like amputation, kidney damage, or neuropathy.

\noindent\textit{Explanation:} Includes significant health effects (e.g., nephropathy, retinopathy) stemming from uncontrolled diabetes. Multiple minor complications combined may also qualify.

\noindent\textbf{MAKES-DECISIONS}

\noindent\textit{Criteria:} Patient must make their own medical decisions.

\noindent\textit{Explanation:} This assesses cognitive ability to understand medical options, communicate choices, and appreciate consequences.

\noindent\textbf{MI-6MOS}

\noindent\textit{Criteria:} Myocardial infarction (MI) within the past 6 months.

\noindent\textit{Explanation:} MI refers to heart muscle damage from blocked blood flow. Assess via medical history and records confirming the event within 6 months.

\subsection{Criteria from ClinicalTrials.gov}\label{sec:clinicalgov_cri}
\noindent\textbf{Inclusion Criteria}

\noindent 1. Locally advanced or metastatic breast cancer confirmed by histopathology.

\noindent 2. Received 2 previous lines of anti-tumor treatment and developed resistance to standard treatment.

\noindent 3. Patients with locally advanced or metastatic advanced solid tumors confirmed by histology or cytology (including but not limited to triple-negative breast cancer, gastric cancer, colorectal cancer); Patients with 1 line of standard treatment failure (disease progression after treatment or intolerability of toxic side effects of treatment), or no standard treatment, or unable to receive standard treatment.

\noindent 4. At least 1 measurable lesion per Response Evaluation Criteria in Solid Tumors version 1.1 and previously treated lesions with radiotherapy or focal therapy and no progression cannot be included as target lesion for assessment.

\noindent 5. Histologically confirmed breast cancer with an invasive component measuring 20 mm and/or with morphologically confirmed spread to regional lymph nodes (stage cT2-cT4 with any cN, or cN1-cN3 with any cT).

\noindent 6. Known HER2-positive breast cancer defined as an IHC status of 3+. If IHC is 2+, a positive in situ hybridization (FISH, CISH, or SISH) test is required by local laboratory testing.

\noindent 7. Patients with breast cancer planned to receive Anthracycline and Cyclophosphamide chemotherapy.

\noindent 8. Patients treated for hormone-dependent localized breast cancer requiring adjuvant hormonal therapy (HT).

\noindent 9. Patients treated with tamoxifen for a maximum of 1 to 3 years.

\noindent 10. Histologically confirmed ER+/HER2- early-stage resected invasive breast cancer at high or intermediate risk of recurrence, based on clinical-pathological risk features, as defined in the protocol.

\noindent 11. Completed adequate (definitive) locoregional therapy (surgery with or without radiotherapy) for the primary breast tumour(s), with or without (neo)adjuvant chemotherapy.

\noindent 12. Completed at least 2 years but no more than 5 years (+3 months) of adjuvant ET (+/- CDK4/6 inhibitor).

\noindent 13. Patients receiving letrozole for more than two months.

\noindent 14. Postmenopausal hormone receptor-positive patients.

\noindent 15. Patients with N0 disease or those who have undergone targeted axillary detection with a good response.

\medskip

\noindent\textbf{Exclusion Criteria}

\noindent 1. Patients known or suspected intolerance or hypersensitivity to main ingredient or any of the excipients of SNB-101.

\noindent 2. Patients with bilateral invasive breast cancers.

\noindent 3. Age $<$18 years old or $>$70 years old.

\noindent 4. Patients with standard metallic contraindications to CMR or estimated glomerular filtration rate $<$30 mL/min/1.73 m².

\noindent 5. History of hypersensitivity or contraindication to TTF.

\noindent 6. Implanted pacemaker, defibrillator, or other electrical medical devices.

\noindent 7. Bilateral breast cancer (including multifocal breast cancer. All histological lesions were HER2 1+ or 2+, not detected by FISH amplification and HR positive according to ASCO guidelines).

\noindent 8. Received any form of anti-tumor therapy (chemotherapy, radiotherapy, molecular targeted therapy, endocrine therapy, etc.).

\noindent 9. Breast cancer without histopathological diagnosis.

\noindent 10. Known personal history of ductal carcinoma in situ (DCIS) or invasive breast cancer.

\noindent 11. Prior systemic treatment for any malignancy.

\noindent 12. Significant medical comorbidities as per investigator evaluation.

\noindent 13. Patients with ongoing indications for the cardioprotective medication - ACE inhibitors, ARBs and/or beta-blockers.

\noindent 14. Intermediate or high-grade ductal carcinoma in situ.

\noindent 15. Invasive carcinoma.

\subsection{Example Criteria from ClinicalTrials.gov refined by GPT-4o}\label{sec:clinicalgov_cri_gpt}
\noindent
\textit{Criterion:} Locally advanced or metastatic breast cancer confirmed by histopathology.\\
\textit{Additional Explanation:}\\
- Locally advanced breast cancer refers to cancer that has spread beyond the breast to nearby areas but not to distant body parts. It often involves large tumors or extensive lymph node involvement. \\
- Metastatic breast cancer means that the cancer has spread from the breast to other parts of the body, such as the bones, liver, lungs, or brain.\\
- Histopathology is the examination of tissues under a microscope to study the manifestations of disease. This confirms the diagnosis of breast cancer by analyzing tissue samples obtained from a biopsy or surgery.\\

\vspace{1em}
\noindent
\textit{Criterion:} Received 2 previous lines of anti-tumor treatment and developed resistance to standard treatment.\\
\textit{Additional Explanation:}\\
1. **Anti-tumor treatment**: This refers to any therapy aimed at treating cancer, which might include chemotherapy, targeted therapy, immunotherapy, or hormone therapy.\\
2. **2 previous lines of treatment**: In cancer therapy, a "line" of treatment refers to a complete sequence of therapy given for cancer, which can consist of one or multiple medications or interventions. The patient must have undergone two different complete treatment sequences.\\
3. **Developed resistance**: This means that the cancer no longer responds to the standard treatment. This can be defined by the tumor growing or not shrinking in size despite treatment, or the cancer markers take not being reduced.\\
4. **Standard treatment**: This refers to the most widely accepted and utilized treatments for a particular type of cancer, as established by clinical research and guidelines. Examples might include first-line chemotherapy regimens or targeted therapies. Exact "standard treatment" could vary based on cancer type, but usually involves those strategies that have the highest success rates based on evidence.\\

\section{Prompts}
\subsection{The prompt for zero-shot matching}\label{app:prompt_zero}
You are an experienced physician tasked with determining whether a patient meets specific eligibility criteria for clinical trials. Below, I will provide multiple EHR records for a single patient along with one criterion. You must evaluate whether the patient satisfies the criterion. For each criterion, respond strictly with 1 if the criterion is met, or 0 if it is not met. If there is not enough information to make a determination, return 0. Respond with only 0 or 1; no reasoning or explanation is needed.\\
Patient EHR Data:...\\
The Ctiterion:...
\subsection{The prompt for refining criteria}\label{app:prompt_gpt_refine}
You are an experienced physician tasked with reviewing the trial criterion to make it easy to follow for less-trained staff. Sometimes, the criterion may seem vague. If it is, please complement the criterion for clarity in plain English by: (1) defining any specialized medical terms, and (2) explaining vague phrases such as ‘normal levels’ by specifying normal ranges for different genders and age groups. You must keep the original criterion intact and add additional explanations only if you think they are necessary. Please ONLY return the revised criterion and use the following format:\\
Original Criterion: XXX \\
Additional Explanation: XXX \\
Here is the original criterion:...

\subsection{Prompts for Agents}\label{app:prompt_agent}
\subsubsection{Router Agent} \label{app:prompt_router}
You are a routing agent responsible for selecting the most appropriate augmentation agent to enrich a trial criterion with relevant, domain-specific information. You will receive one criterion as input. Based on its content, determine which augmentation agent to use.\\
**Retrieval Agent**: Use this agent if the criterion requires authoritative information from indexed medical databases like MedDRA and FDALabel.\\
**Online Search Agent**: Use this agent if the criterion needs supplementary knowledge from external online sources.\\
**Self-Augment Agent**: Use this agent if the criterion can be refined using the language model’s intrinsic medical knowledge.\\
Strictly respond with only one of the following labels indicating your choice: Retrieval, Online Search, or Self-Augment. No reasoning or explanation is needed—only return the label.\\
The Criterion:..
\subsubsection{Knowledge Augmentation Agent} \label{app: prompt_aug}
You are an experienced physician tasked with reviewing the trial criterion to make it easy to follow for less-trained staff. Sometimes, the criterion may seem vague. If it is, please complement the criterion for clarity in plain English by: (1) defining any specialized medical terms, and (2) explaining vague phrases such as ‘normal levels’ by specifying normal ranges for different genders and age groups. You must keep the original criterion intact and add additional explanations only if you think they are necessary. Please ONLY return the revised criterion and use the following format:\\
Original Criterion: XXX \\
Additional Explanation: XXX \\
Here is the original criterion:...
\subsubsection{Critic Agent} \label{app:prompt_critic}
You are a senior physician reviewing criteria revised by another physician. Make sure that (1) the original criterion is preserved, and (2) any additional explanations are accurate. For each criterion, respond strictly with **1** if both checks pass, or **0** if they do not. No reasoning or explanation is needed.\\
The Revised Ctiterion:...\\
The Original Ctiterion:...
\subsubsection{Reasoning Agent} \label{app:prompt_reason}
You are a Principal Investigator (PI) in a clinical trial. You have multiple EHR records for one patient and a criterion with detailed explanation. Please perform a reasoning process to analyze why this patient meets or does not meet this trial criterion. If the information is not sufficient to make a determination, assume the patient does not meet the criterion.\\
Patient EHR Data:...\\
The Detailed Ctiterion:...
\subsubsection{Matching Agent} \label{app:prompt_matching}
You are a Principal Investigator (PI) in a clinical trial and are tasked with determining whether a patient meets specific eligibility criteria. Below, I will provide multiple EHR records for one patient, one criterion, and a reasoning process from another PI. You must evaluate whether the patient meets the criterion. For each criterion, respond strictly with **1** if met or **0** if not met. Respond with 0 or 1 only; no reasoning or explanation is needed.\\
Patient EHR Data:...\\
The Ctiterion:...\\
The Reasoning Process:...

\subsection{Pseudo-Code for MAKAR Workflow}\label{app:pseudio}
\input{table/persudocode}

\end{document}

%% file: table/main_result.tex
\begin{table*}[t]
\centering

\caption{Performance Comparison Across Models on N2C2 and ClinicalTrial Datasets.}
\label{tab:main_result}
\small
\begin{tabularx}{\textwidth}{
  p{0.15\textwidth}
  *{8}{>{\centering\arraybackslash}X}
}
\toprule
\textbf{Criteria} &
\multicolumn{2}{c}{TrialGPT~\cite{jin2024matching}} &
\multicolumn{2}{c}{Wornow et al.~\cite{wornow2024zero}} &
\multicolumn{2}{c}{Beattie et al.~\cite{beattie2024utilizing}} &
\multicolumn{2}{c}{\textbf{MAKAR}} \\
\cmidrule(lr){2-3}
\cmidrule(lr){4-5}
\cmidrule(lr){6-7}
\cmidrule(lr){8-9}
& {Acc.} & {F1} & {Acc.} & {F1} & {Acc.} & {F1} & {Acc.} & {F1} \\
\midrule
\multicolumn{9}{c}{\textbf{N2C2 Dataset}} \\
\midrule
ABDOMINAL       & 0.813 & 0.722 & 0.858 & 0.785 & \textbf{0.868} & \textbf{0.804} & 0.809 & 0.671 \\
ADVANCED-CAD    & 0.823 & 0.859 & 0.854 & 0.873 & 0.854 & 0.877 & \textbf{0.882} & \textbf{0.897} \\
ALCOHOL-ABUSE   & 0.972 & 0.667 & 0.969 & 0.667 & 0.969 & 0.571 & \textbf{0.983} & \textbf{0.706} \\
ASP-FOR-MI      & \textbf{0.906} & \textbf{0.944} & 0.861 & 0.915 & 0.875 & 0.923 & 0.896 & 0.938 \\
CREATININE      & 0.802 & 0.776 & 0.802 & 0.776 & 0.802 & 0.776 & \textbf{0.924} & \textbf{0.900} \\
DIETSUPP-2MOS   & 0.722 & 0.759 & 0.764 & 0.805 & 0.781 & 0.818 & \textbf{0.826} & \textbf{0.846} \\
DRUG-ABUSE      & 0.882 & 0.469 & 0.941 & 0.638 & 0.920 & 0.566 & \textbf{0.972} & \textbf{0.765} \\
ENGLISH         & 0.948 & 0.971 & 0.976 & 0.987 & 0.986 & 0.992 & \textbf{0.997} & \textbf{0.998} \\
HBA1C           & 0.840 & 0.800 & 0.840 & 0.798 & 0.830 & 0.790 & \textbf{0.934} & \textbf{0.899} \\
KETO-1YR        & \textbf{0.993} & {-} & \textbf{0.993} & {-} & 0.990 & {-} & \textbf{0.993} & {-} \\
MAJOR-DIABETES  & 0.830 & 0.838 & 0.833 & \textbf{0.845} & 0.826 & 0.842 & \textbf{0.840} & 0.838 \\
MAKES-DECISIONS & 0.580 & 0.722 & 0.920 & 0.957 & 0.924 & 0.959 & \textbf{0.965} & \textbf{0.982} \\
MI-6MOS         & 0.833 & 0.510 & 0.885 & 0.582 & 0.872 & 0.532 & \textbf{0.962} & \textbf{0.807} \\
\midrule
\textbf{Average} & 0.842 & 0.753 & 0.884 & 0.802 & 0.884 & 0.787 & \textbf{0.922} & \textbf{0.854} \\
\midrule
\multicolumn{9}{c}{\textbf{ClinicalTrial Dataset}} \\
\midrule
CT-IN & 0.900 & 0.828 & {-} & {-} & 0.967 & 0.929 & \textbf{0.987} & \textbf{0.982} \\
CT-EX & 0.900 & 0.885 & {-} & {-} & 0.947 & 0.937 & \textbf{0.987} & \textbf{0.982} \\
\bottomrule
\\
\multicolumn{9}{p{\textwidth}}{\raggedright\footnotesize
 $^{\mathrm{a}}$ CT-IN and CT-EX denote inclusion and exclusion criteria in the ClinicalTrial dataset. The F1 score for KETO-1YR is unavailable because only one patient met this criterion, and no method matched it correctly.}
\end{tabularx}
\end{table*}

%% file: table/ablation1.tex
% \begin{table}[ht]
% \centering
% \normalsize
% \caption{Results of Ablation Analysis. }
% \resizebox{\columnwidth}{!}{
% \begin{tabular}{
% l
% S[table-format=1.3]
% S[table-format=1.3]
% S[table-format=1.3]
% S[table-format=1.3]
% }
% \toprule
% \textbf{Ablations} & \textbf{Zero Shot} & \textbf{MAKAR-Reason} & \textbf{MAKAR-Aug} & \textbf{MAKAR} \\
% \midrule
% F1-Score                  &0.857 & 0.928 & 0.933 & 0.982 \\
% \midrule
% \multicolumn{5}{l}{\textbf{Reasoning Models}} \\
% OpenAI o1                     & 0.898 & {-} & {-} & {-} \\
% OpenAI o3 mini                & 0.825 & {-} & {-} & {-} \\

% \bottomrule
% \hline
% \multicolumn{4}{l}{$^{\mathrm{a}}$We report only zero-shot results for reasoning models here, as they are not suited for CoT strategies \cite{guo2025deepseek}.}
% \end{tabular}
% }

% \label{tab:ablation1}
% \end{table}
\begin{table}[t]
\centering
\footnotesize
\caption{Results of Ablation Analysis.}
\label{tab:ablation1}
\begin{tabularx}{\columnwidth}{
    >{\raggedright\arraybackslash}X
    *{4}{>{\centering\arraybackslash}X}
}
\toprule
\textbf{Ablations} &
\textbf{Zero Shot} &
\textbf{MAKAR-Reason} &
\textbf{MAKAR-Aug} &
\textbf{MAKAR} \\
\midrule
F1-Score & 0.857 & 0.928 & 0.933 & 0.982 \\
\midrule
\multicolumn{5}{l}{\textbf{Reasoning Models}} \\
OpenAI o1      & 0.898 & -- & -- & -- \\
\mbox{OpenAI o3 mini} & 0.825 & -- & -- & -- \\
\bottomrule
\multicolumn{5}{p{\columnwidth}}{\footnotesize 
$^{\mathrm{a}}$We report only zero-shot results for reasoning models here, as they are not suited for CoT strategies \cite{guo2025deepseek}.}
\end{tabularx}
\end{table}

%% file: table/ablation2.tex
\begin{table}[ht]
\centering
\normalsize
\caption{Results of Model Ablation in the Reasoning Module.}
\resizebox{\columnwidth}{!}{
\begin{tabular}{
l
S[table-format=1.3]
S[table-format=1.3]
S[table-format=1.3]
}
\toprule
\textbf{Underlying Models} & \textbf{Zero Shot} & \textbf{GPT-Aug} & \textbf{MAKAR} \\
\midrule
Qwen 2.5 Instruct 7B      & 0.853 & 0.853 & 0.900 \\
Qwen 2.5 Instruct 14B     & 0.869 & 0.900 & 0.915 \\
LLaMA 3.1 Instruct 7B     & 0.815 & 0.815 & 0.853 \\
Gemma 3 Instruct 4B       & 0.788 & 0.788 & 0.828 \\
Gemma 3 Instruct 12B      & 0.860 & 0.869 & 0.884 \\
Mistral Instruct v0.3 7B  & 0.770 & 0.770 & 0.870 \\
GPT4o                     & \textbf{0.900} & \textbf{0.943} & \textbf{0.987} \\             
\bottomrule
\hline
\multicolumn{4}{l}{$^{\mathrm{a}}$ GPT-Aug based on GPT-4o is equivalent to MAKAR-Aug.}
\end{tabular}
}
\label{tab:ablation2}
\end{table}

%% file: table/stats.tex
\begin{table}[h]
    \centering
    \caption{Counts of Eligible and Ineligible Patients in N2C2}
    \small
    \begin{tabular}{lcc}
        \toprule
        \textbf{Criterion} & \textbf{Met Count} & \textbf{Not Met Count} \\
        \midrule
        ABDOMINAL          & 107 & 181 \\
        ADVANCED-CAD       & 170 & 118 \\
        ALCOHOL-ABUSE      & 10  & 278 \\
        ASP-FOR-MI         & 230 & 58  \\
        CREATININE         & 106 & 182 \\
        DIETSUPP-2MOS      & 149 & 139 \\
        DRUG-ABUSE         & 15  & 273 \\
        ENGLISH            & 265 & 23  \\
        HBA1C              & 102 & 186 \\
        KETO-1YR           & 1   & 287 \\
        MAJOR-DIABETES     & 156 & 132 \\
        MAKES-DECISIONS    & 277 & 11  \\
        MI-6MOS            & 26  & 262 \\
        \bottomrule
    \end{tabular}
    
    \label{tab:met_counts}
\end{table}

%% file: table/persudocode.tex
\begin{algorithm}
\caption{MAKAR: Multi-Agent Knowledge Augmentation and Reasoning}
\label{app:alg}
\begin{algorithmic}[1]
\Require Original criterion $C_{orig}$, Patient EHR data $P_{EHR}$
\Ensure Matching result $M \in \{\text{0}, \text{1}\}$

\State \textbf{// Augmentation Module: MAKAR-Aug}
\State $C_{aug} \gets \text{RouterAgent}(C_{orig})$
\If{$\text{RouterAgent determines } C_{orig} \text{ is detailed enough}$}
    \State $C_{aug} \gets C_{orig}$ \Comment{Forward directly to Reasoning Module}
\Else
    \State $\text{agent\_type} \gets \text{RouterAgent.selectAgent}(C_{orig})$
    \If{$\text{agent\_type} = \text{retrieval}$}
        \State $C_{aug} \gets \text{RetrievalAgent}(C_{orig}, \text{databases})$
    \ElsIf{$\text{agent\_type} = \text{online\_search}$}
        \State $C_{aug} \gets \text{OnlineSearchAgent}(C_{orig})$
    \ElsIf{$\text{agent\_type} = \text{self\_augment}$}
        \State $C_{aug} \gets \text{SelfAugmentAgent}(C_{orig})$
    \EndIf
    
    \State \textbf{// Iterative refinement with Critic Agent}
    \Repeat
        \State $\text{feedback} \gets \text{CriticAgent}(C_{aug}, C_{orig})$
        \If{$\text{feedback.rejected}$}
            \State $C_{aug} \gets \text{Re-augment}(C_{orig}, \text{feedback})$
        \EndIf
    \Until{$\text{CriticAgent accepts } C_{aug}$}
\EndIf

\State \textbf{// Reasoning Module: MAKAR-Reason}
\State $\text{reasoning\_summary} \gets \text{ReasoningAgent}(C_{aug}, P_{EHR})$

\State \textbf{// Generate detailed evaluation}
\For{each concept $c$ in $C_{aug}$}
    \State $\text{evaluation}[c] \gets \text{ReasoningAgent.evaluate}(c, P_{EHR})$
\EndFor

\State \textbf{// Final matching decision}
\State $M \gets \text{MatchingAgent}(C_{aug}, P_{EHR}, \text{reasoning\_summary})$

\State \textbf{// Supervision validation}
\State $\text{validation} \gets \text{SupervisionMatchingAgent}(M, C_{aug}, P_{EHR})$
\If{$\text{validation.approved}$}
    \State \Return $M$
\Else
    \State \Return $\text{validation.corrected\_result}$
\EndIf

\end{algorithmic}
\end{algorithm}